\documentclass[twoside,twocolumn,9pt]{article}
\usepackage{extsizes}
\usepackage[super,sort&compress,comma]{natbib} 
\usepackage[version=3]{mhchem}
\usepackage[left=1.5cm, right=1.5cm, top=1.785cm, bottom=2.0cm]{geometry}
\usepackage{balance}
\usepackage{mathptmx}
\usepackage{sectsty}
\usepackage{graphicx} 
\usepackage{lastpage}
\usepackage[format=plain,justification=justified,singlelinecheck=false,font={stretch=1.125,small,sf},labelfont=bf,labelsep=space]{caption}
\usepackage{float}
\usepackage{fancyhdr}
\usepackage{fnpos}
\usepackage[english]{babel}
\addto{\captionsenglish}{%
  
}
\usepackage{array}
\usepackage{droidsans}
\usepackage{charter}
\usepackage[T1]{fontenc}
\usepackage[usenames,dvipsnames]{xcolor}
\usepackage{setspace}
\usepackage[compact]{titlesec}
\usepackage{hyperref}

\usepackage{amsmath}
\numberwithin{equation}{section}
\usepackage{amssymb}								
\usepackage{graphicx}								 	

\usepackage{epstopdf}

\definecolor{cream}{RGB}{222,217,201}

\begin{document}

\pagestyle{fancy}
\thispagestyle{plain}
\fancypagestyle{plain}{
\renewcommand{\headrulewidth}{0pt}
}

\makeFNbottom
\makeatletter
\renewcommand\LARGE{\@setfontsize\LARGE{15pt}{17}}
\renewcommand\Large{\@setfontsize\Large{12pt}{14}}
\renewcommand\large{\@setfontsize\large{10pt}{12}}
\renewcommand\footnotesize{\@setfontsize\footnotesize{7pt}{10}}
\makeatother

\renewcommand{\thefootnote}{\fnsymbol{footnote}}
\renewcommand\footnoterule{\vspace*{1pt}%
\color{cream}\hrule width 3.5in height 0.4pt \color{black}\vspace*{5pt}} 
\setcounter{secnumdepth}{5}
\newcommand*{\dc}{DC$_{50}$}
\newcommand*{\dmax}{D$_{\text{max}}$}
\makeatletter 
\renewcommand\@biblabel[1]{#1}            
\renewcommand\@makefntext[1]%
{\noindent\makebox[0pt][r]{\@thefnmark\,}#1}
\makeatother 
\renewcommand{\figurename}{\small{Fig.}~}
\sectionfont{\sffamily\Large}
\subsectionfont{\normalsize}
\subsubsectionfont{\bf}
\setstretch{1.125} 
\setlength{\skip\footins}{0.8cm}
\setlength{\footnotesep}{0.25cm}
\setlength{\jot}{10pt}
\titlespacing*{\section}{0pt}{4pt}{4pt}
\titlespacing*{\subsection}{0pt}{15pt}{1pt}

\fancyfoot{}
\fancyfoot[LO,RE]{\vspace{-7.1pt}\includegraphics[height=9pt]{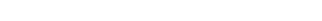}}
\fancyfoot[CO]{\vspace{-7.1pt}\hspace{13.2cm}\includegraphics{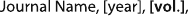}}
\fancyfoot[CE]{\vspace{-7.2pt}\hspace{-14.2cm}\includegraphics{headfoot/RF}}
\fancyfoot[RO]{\footnotesize{\sffamily{1--\pageref{LastPage} ~\textbar  \hspace{2pt}\thepage}}}
\fancyfoot[LE]{\footnotesize{\sffamily{\thepage~\textbar\hspace{3.45cm} 1--\pageref{LastPage}}}}
\fancyhead{}
\renewcommand{\headrulewidth}{0pt} 
\renewcommand{\footrulewidth}{0pt}
\setlength{\arrayrulewidth}{1pt}
\setlength{\columnsep}{6.5mm}
\setlength\bibsep{1pt}

\makeatletter 
\newlength{\figrulesep} 
\setlength{\figrulesep}{0.5\textfloatsep} 

\newcommand{\topfigrule}{\vspace*{-1pt}%
\noindent{\color{cream}\rule[-\figrulesep]{\columnwidth}{1.5pt}} }

\newcommand{\botfigrule}{\vspace*{-2pt}%
\noindent{\color{cream}\rule[\figrulesep]{\columnwidth}{1.5pt}} }

\newcommand{\dblfigrule}{\vspace*{-1pt}%
\noindent{\color{cream}\rule[-\figrulesep]{\textwidth}{1.5pt}} }

\makeatother

\twocolumn[
  \begin{@twocolumnfalse}
{\includegraphics[height=30pt]{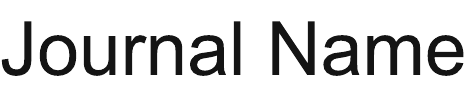}\hfill\raisebox{0pt}[0pt][0pt]{\includegraphics[height=55pt]{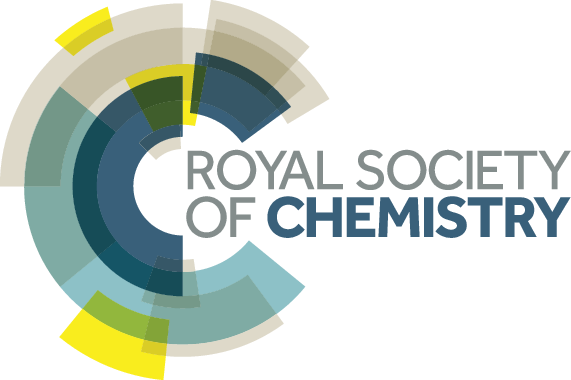}}\\[1ex]
\includegraphics[width=18.5cm]{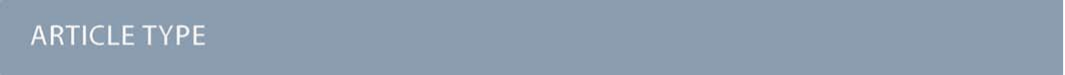}}\par
\vspace{1em}
\sffamily
\begin{tabular}{m{4.5cm} p{13.5cm} }
\includegraphics{
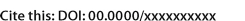} & \noindent\LARGE{\textbf{A Comprehensive Review of Emerging Approaches in Machine Learning for \textit{De Novo} PROTAC Design 
}} \\
\vspace{0.3cm} & \vspace{0.3cm} \\

 & \noindent\large{Yossra Gharbi and Rocío Mercado} \\


\includegraphics{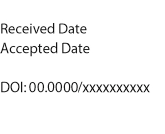} & \noindent\normalsize{Targeted protein degradation (TPD) is a rapidly growing field in modern drug discovery that aims to regulate the intracellular levels of proteins by harnessing the cell's innate degradation pathways to selectively target and degrade disease-related proteins. This strategy creates new opportunities for therapeutic intervention in cases where occupancy-based inhibitors have not been successful. Proteolysis-targeting chimeras (PROTACs) are at the heart of TPD strategies, which leverage the ubiquitin-proteasome system for the selective targeting and proteasomal degradation of pathogenic proteins. This unique mechanism can be particularly useful for dealing with proteins that were once deemed ``undruggable'' using conventional small-molecule drugs. PROTACs are hetero-bifunctional molecules consisting of two ligands, connected by a chemical linker. As the field evolves, it becomes increasingly apparent that the traditional methodologies for designing such complex molecules have limitations. This has led to the use of machine learning (ML) and generative modeling to improve and accelerate the development process. In this review, we aim to provide a thorough exploration of the impact of ML on \textit{de novo} PROTAC design — an aspect of molecular design that has not been comprehensively reviewed despite its significance. Initially, we delve into the distinct characteristics of PROTAC linker design, underscoring the complexities required to create effective bifunctional molecules capable of TPD. We then examine how ML in the context of fragment-based drug design (FBDD), honed in the realm of small-molecule drug discovery, is paving the way for PROTAC linker design. Our review provides a critical evaluation of the limitations inherent in applying this method to the complex field of PROTAC development.  Moreover, we review existing ML works applied to PROTAC design, highlighting pioneering efforts and, importantly, the limitations these studies face. By offering insights into the current state of PROTAC development and the integral role of ML in PROTAC design, we aim to provide valuable perspectives for biologists, chemists, and ML practitioners alike in their pursuit of better design strategies for this new modality.
}

\end{tabular}

 \end{@twocolumnfalse} \vspace{0.6cm}

  ]

\renewcommand*\rmdefault{bch}\normalfont\upshape
\rmfamily
\section*{}
\vspace{-1cm}


\footnotetext{\textit{Department of Computer Science and Engineering, Section for Data Science and AI, Chalmers University of Technology, Chalmersplatsen 4, 412 96 Gothenburg, Sweden.  E-mail: yossra@chalmers.se; rocio.mercado@chalmers.se.}}


\section{Introduction}

\begin{figure*}[h!]
 \centering
 \includegraphics[scale=0.6]{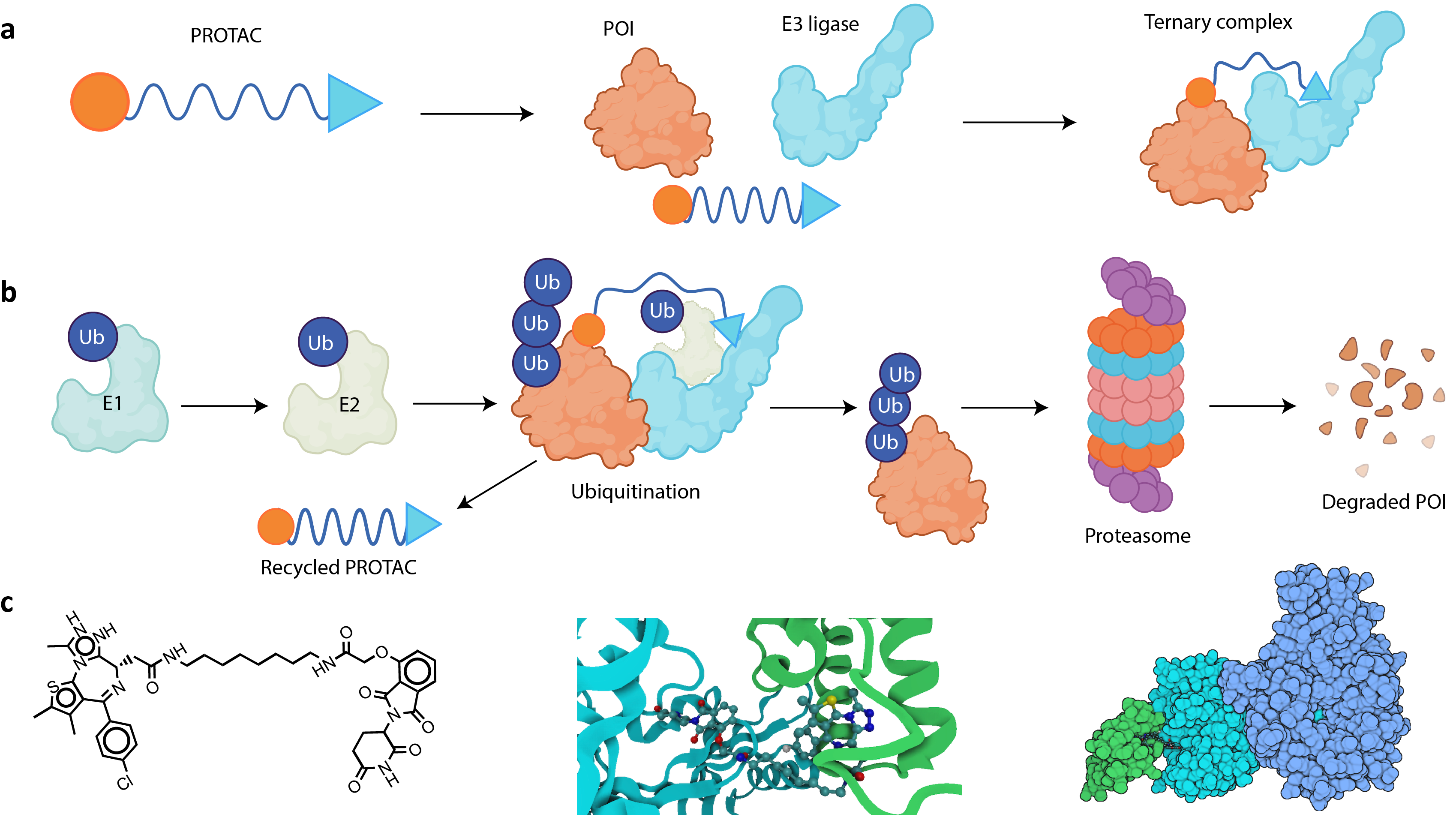}
 \caption{(a) A PROTAC is a hetero-bifunctional molecule, consisting of a ligand (blue triangle) that recruits an E3 ubiquitin ligase, a warhead (orange circle) that binds to the POI, and a linker (blue curve) that connects the two binding moieties. The PROTAC functions by simultaneously binding to the POI and the E3 ligase, thus bringing them into close proximity and inducing the formation of a ternary complex. (b) The PROTAC MoA begins with an E1 ubiquitin-activating enzyme that activates ubiquitin (Ub) in an ATP-dependent manner. This activated Ub is then transferred to an E2 Ub-conjugating enzyme. Subsequently, a PROTAC simultaneously binds to the POI and an E3 ubiquitin ligase, bringing them into close proximity. This facilitates the transfer of Ub from the E2 enzyme to the POI, catalyzed by the E3 ligase. The polyubiquitinated POI is then recognized and degraded by the proteasome into smaller peptides, and the PROTAC is released back into the cellular environment where it can be reused, initiating the process again with another instance of the same POI. (c) Visual representations of dBET6 and its respective ternary complex: \textit{left} -- a 2D skeletal formula of the PROTAC molecule dBET6; \textit{middle} -- a close-up of the dBET6 degrader's three-dimensional (3D) structure in complex with CRBN and BRD4 (PDBID:6BOY), emphasizing the importance of the PROTAC's spatial orientation in forming a good ternary complex; and \textit{right} -- a space filling model for the same complex, involving BRD4, CRBN, DNA damage-binding protein 1 (DDB1), and dBET6. Color key: BRD4 (green), CRBN (cyan), and DDB1 (dark blue). }.
 \label{fgr:MoA}
\end{figure*}

Targeted protein degradation (TPD) is a novel therapeutic approach with attractive potential to eliminate disease-causing proteins from within cells.\cite{lai2017induced, schapira2019targeted, Békés2022PROTAC, zhao2022targeted} Traditional drug development strategies have focused on inhibiting the activity of such proteins, but TPD goes a step further by removing or reducing protein levels from the cell. 
This is particularly useful for targeting proteins that are difficult to inhibit with small molecules or biologics, often due to the absence of well-defined binding sites; these are frequently referred to as ``undruggable'' targets, and they can be challenging to target due to their structure, location, and/or function. \cite{kim2014privileged, martin2021protacs} Proteolysis-targeting chimeras (PROTACs) are hetero-bifunctional molecules engineered to bind simultaneously to an E3 ligase, a key enzyme involved in the process of tagging proteins for degradation\cite{clague2015demographics}, and the protein of interest (POI) that is targeted for degradation (Figure \ref{fgr:MoA}a). \cite{gu2018protacs, xi2019small, liu2022overview} A PROTAC molecule brings the E3 ligase and the POI into close proximity, facilitating the formation of a ternary complex consisting of the E3 ligase system, the PROTAC, and the POI to induce POI ubiquitination and its subsequent degradation by the proteasome (Figure \ref{fgr:MoA}b). \cite{coleman2018proteolysis}

PROTACs were first reported in 2001, when the first fully synthesized PROTAC, named Protac-1, was developed by Crews, Deshaies, and co-workers.\cite{sakamoto2001protacs}
This provided an \textit{in vitro} proof-of-concept, which proved the feasibility of designing molecules with the potential to selectively target and degrade cellular proteins by hijacking the ubiquitin proteasome system (UPS). Protac-1 was specifically designed to target the methionyl aminopeptidase 2 (MetAP-2) protein, which plays a role in angiogenesis and various other pathologies, including cancer. Protac-1 was designed to target MetAP-2, as ovalicin and fumagillin do, but with the added mechanism of promoting its degradation. The binding of Protac-1 to MetAP-2 led to the tethering of MetAP-2 to beta-transducin repeat-containing protein ($\beta$-TrCP), functioning as an E3 ubiquitin ligase responsible for ubiquitination of MetAP-2. The effectiveness of Protac-1 in facilitating the ubiquitination of MetAP-2 was demonstrated using extracts from unfertilized \textit{Xenopus laevis} eggs, a common model organism in biomedical research. \cite{cannatella1993xenopus} These extracts provided a controlled environment rich in cellular machinery that mimics the conditions inside a living cell needed for ubiquitination, protein degradation, and observing the interaction between Protac-1 and MetAP-2. Results showed that MetAP-2 selectively binds the angiogenesis inhibitor ovalicin moiety of Protac-1, and that Protac-1 can mediate the ubiquitination of MetAP-2 by $\beta$-TrCP, leading to its degradation. \cite{sakamoto2001protacs}

In 2003, the same group synthesized a PROTAC using estradiol, a form of estrogen, as part of its structure.\cite{sakamoto2003development} This PROTAC was designed to target and promote the destruction of the estrogen receptor alpha (ER$ \alpha$), which, when activated by estrogen, can promote the growth of some breast cancers.\cite{renoir2013estrogen} It has been shown that the estradiol-based PROTAC could effectively enforce the ubiquitination and subsequent degradation of the $\alpha$ isoform of ER \textit{in vitro}. \cite{sakamoto2003development} Similarly, they created a PROTAC that incorporates dihydrotestosterone (DHT) to target and degrade the androgen receptor (AR). When activated by androgens like DHT, the AR can stimulate the growth of prostate cancer cells.\cite{article} The DHT-based PROTAC has shown efficacy in promoting the rapid ubiquitination and proteasome-dependent degradation of AR in cellular tests. \cite{sakamoto2003development}
These PROTACs served as proof that they are a promising modality for selectively degrading key proteins involved in cancer, opening up potential treatment benefits by TPD in hormone-responsive cancers.\cite{bhole2023unlocking, li2020proteolysis} 

While first-generation PROTACs were capable of degrading target proteins, they suffered from poor cell permeability and chemical stability stemming from their high molecular weight. \cite{liu2020protacs} They generally exhibited low potency using micromolar (µM) concentrations, which is less desirable than the nanomolar concentrations used with more potent drugs, indicating that higher doses are required to exhibit efficacy. \cite{lai2017induced} Notably, early PROTACs were peptide-based and commonly used $\beta$-TrCP or Von Hippel-Lindau (VHL) as E3 ligases. One significant drawback of peptide-based therapeutics is their high molecular weight, which affects their ability to cross cell membranes. This poor permeability is a critical limitation because it means that even if a PROTAC is theoretically effective, its inability to enter cells renders them ineffective in practice. \cite{lai2017induced}
These limitations promoted the need to develop second-generation PROTACs, motivating a transition from peptide-based to small-molecule PROTACs. The use of small molecules expanded the range of potentially targetable proteins by taking advantage of a more extensive array of E3 ligases beyond $\beta$-TrCP and VHL, such as mouse double minute 2 homologue (MDM2), inhibitors of apoptosis proteins (IAPs), and cereblon (CRBN). \cite{liu2020protacs} In 2008, the Crews lab developed the first small-molecule PROTAC that could degrade a target protein within cells, in this case, targeting AR. \cite{schneekloth2008targeted} This PROTAC was composed of nutlin-3A, a ligand for MDM2, and a non-steroidal androgen receptor ligand (SARM) for AR, connected by a polyethylene glycol (PEG) linker.\cite{schneekloth2008targeted} The SARM-nutlin PROTAC induced the degradation of AR in a proteasome-dependent manner with enhanced cell penetration \textit{in vitro}.

In some of the latest generations of PROTACs, additional elements have been introduced to give another dimension of control over PROTAC activity. \cite{liu2021light, xiao2022recent} These classes of controllable PROTACs aim to address off-tissue effects by controlling PROTAC action in a spatiotemporal manner. \cite{xiao2022recent} Some are designed to be activated or deactivated by specific wavelengths of light, allowing for controlled degradation processes in target cells, with potentially reduced side effects and enhanced therapeutic index. These PROTACs include phospho-dependent PROTACs that degrade targets with activated kinase-signaling clues, and light-controllable PROTACs that use light as an external clue to trigger target degradation. Notable light-controllable PROTACs, also commonly referred to as PHOTACs, include photo-caged and photo-switchable PROTACs.\cite{reynders2020photacs} Photo-caged PROTACs are designed to be inactive in their initial form and activated by light exposure, which removes the photo-cage group and enables the degradation of the POI. Photo-switchable PROTACs, on the other hand, are designed to reversibly control the degradation process via the incorporation of photoswitchable groups such as azobenzene, which can switch between active and inactive states under different wavelengths of light.
In-cell click-formed proteolysis-targeting chimeras (CLIPTACs) share similar ambitions to PHOTACs and have been used to degrade two key oncology targets successfully.\cite{lebraud2016protein}
The reader is referred to these excellent reviews for a more detailed analysis of milestones in PROTAC development.\cite{lai2017induced, pettersson2019proteolysis, liu2020protacs, xiao2022recent}

Since the first PROTAC was reported in the literature, the field of PROTACs has experienced remarkable growth\cite{li2021protein, Békés2022PROTAC} and has led to the design of compounds with improved drug-like properties demonstrating effectiveness both \textit{in vitro} and \textit{in vivo}.\cite{ma2013targeted, bai2019potent, wang2021vitro, dragovich2021antibody} The first PROTACs to enter clinical trials were ARV-110\cite{neklesa2019arv, petrylak2020first} and ARV-471\cite{snyder2021discovery}, which target AR and ER, respectively. ARV-110 was tested in a heavily pre-treated population with metastatic castration-resistant prostate cancer (mCRPC).  Results from a phase I trial showed that ARV-110 could reduce the levels of AR in cancer cells by at least 95\%, which is a significant reduction that hampers the cancer cell's ability to grow and survive.
Notably, its effectiveness in ENZ-resistant models offers a potential treatment option for patients who no longer respond to ENZ, addressing a critical gap in prostate cancer therapy. ARV-110 advanced to phase II clinical trials in 2020 based on initial phase I data that demonstrated the drug's good oral availability, safety, and tolerability in patients\cite{burslem2020proteolysis}.
On the other hand, ARV-471 is designed for oral administration in patients with hormone receptor-positive (HR+) and HER2-negative metastatic breast cancer. In a phase I clinical study involving breast cancer patients who had undergone multiple prior treatments, ARV-471 significantly reduced the expression level of ER in tumor tissues of patients. 
It was also reported that ARV-471 is well tolerated across all tested doses (30-700 mg), and maintained a high level of ER degradation (89\%). \cite{liu2022overview}

Following the lead of ARV-110, thousands of PROTACs have been developed to degrade kinases, nuclear receptors, transcription factors, regulatory proteins, etc.  Some of the most common targets include the bromodomain and extra-terminal domain (BET) family, AR, ER, bruton's tyrosine kinase (BTK), anaplastic lymphoma kinase (ALK), mitogen-activated protein kinase kinase (MEK), the BCR-ABL fusion gene, and the epidermal growth factor receptor (EGFR).\cite{xiao2022recent}. Most of these are well-known targets for cancer and inflammatory disorders. PROTACs are developing rapidly and have already been used to target and degrade a variety of proteins associated with a wide range of diseases, including cancer, immune disorders, neurodegenerative diseases, cardiovascular conditions, and viral infections. \cite{yao2022recent} This rapid progress highlights the utility of PROTACs in treating a wide range of health conditions via TPD.

PROTACs work by ingeniously harnessing the ubiquitin-proteasome system (UPS), an important cellular pathway, which naturally degrades over 80\% of cellular proteins to regulate protein levels and turnover.\cite{chen2016ubiquitin} The UPS selectively targets misfolded and damaged proteins within the cell for degradation, maintaining proper protein homeostasis. \cite{jana2012protein} However, if this system falters such that old, damaged, or surplus proteins are not promptly degraded, they can form aggregates resistant to degradation. These aggregates can interfere with cellular functions and are the hallmark of several neurodegenerative diseases such as Alzheimer's, Parkinson's, and Huntington's diseases.\cite{dantuma2014ubiquitin, schmidt2021ubiquitin, liu2022deubiquitinating, liang2023role} Moreover, a recent multi-omics study of $>$9,000 human tumors and 33 cancer types found that $>$19\% of all cancer driver genes impact UPS function.\cite{tokheim2021systematic} The UPS operates by tagging target proteins for degradation through the attachment of ubiquitin (Ub) protein chains. These Ub tags mark the protein for degradation by the proteasome, a protein complex responsible for protein degradation via proteolysis.\cite{bard2018structure} The process of Ub conjugation involves an enzyme cascade, starting with E1 activating enzymes, proceeding to E2 conjugating enzymes, and culminating with E3 ligases.\cite{lorenz2013macromolecular} Initially, Ub is activated by an E1 enzyme, a reaction that requires adenosine triphosphate and results in an E1-Ub conjugate. The activated Ub is then transferred to an E2 enzyme through a transthioesterification reaction, forming an E2-Ub complex. The most crucial step is mediated by the E3 ubiquitin ligase, which confers specificity to the ubiquitination process. \cite{yang2021e3} 

E3 ubiquitin ligases are categorized into two main types based on their mechanism of action (MoA) for transferring Ub's to their target proteins: HECT-domain and RING-type E3 ligases. HECT-domain E3 ligases first form a thioester bond with Ub. This means that Ub is temporarily attached to the E3 ligase itself. Subsequently, the E3 ligase transfers the Ub from itself directly onto the substrate protein that is to be tagged for degradation. \cite{weber2019hect} Unlike HECT-domain ligases, RING E3 ligases do not form a direct bond with Ub. Instead, they facilitate the transfer of Ub directly from an E2 enzyme (which is conjugated with Ub) to the substrate protein. \cite{branigan2020ubiquitin} In essence, RING-type E3 ligases act as mediators that bring the E2–Ub conjugate close to the substrate, enabling the direct transfer of Ub.
This ubiquitination cycle repeats, leading to the transfer of multiple Ub's and the polyubiquitination of the substrate. Once a protein is polyubiquitinated, it is tagged for degradation. The proteasome recognizes the tagged protein, binds to it, unfolds it, and breaks it down into smaller peptides.\cite{collins2017logic} The PROTAC is then recycled for additional ubiquitination rounds of additional substrates. \cite{sincere2023protacs}

The UPS's sophisticated mechanism forms the basis for PROTAC structure design. PROTACs are bifunctional molecules, designed to harness the UPS for TPD. Each PROTAC comprises three key components: a ligand that binds to the POI, also frequently referred to as the ``warhead;'' a ligand that recruits an E3 ubiquitin ligase; and an organic linker that connects these two ligands. This dual engagement enables PROTACs to bring the POI and E3 ligase into proximity, forming a ternary complex that facilitates the transfer of Ub from the E3 ligase to the target protein. \cite{coleman2018proteolysis, gadd2017structural, wurz2023affinity} Notably, the linker connecting these moieties is not merely a passive scaffold; it has been shown to play a vital role in determining the overall efficacy and specificity of the PROTAC molecule. \cite{zorba2018delineating, chan2018impact, nowak2018plasticity, smith2019differential, guenette2022target} It ensures efficient ubiquitination by correctly positioning the two ligands. This can be achieved by carefully designing the linker length and composition to maintain the required distance, flexibility (or rigidity), and spatial orientation between the POI and the E3 ligase. \cite{troup2020current, zagidullin2020novel, liu2022overview}
Additionally, linker modification can affect properties like hydrogen bond donors (HBDs) and acceptors (HBAs), lipophilicity, molecular weight, rotatable bonds, and polar surface area, which are all critical factors in absorption, distribution, metabolism, and excretion (ADME). \cite{yokoo2023investigating} Improving these properties can make PROTACs function better as drugs. A study on BET degraders provides a notable example,\cite{klein_amide--ester_2021} where researchers replaced an amide bond in BET degraders MZ1 and ARV-771 with an ester group. This change removed one HBD and increased the lipophilicity for each molecule, leading to increased cell membrane permeability. This was reflected in improvements in parallel artificial membrane permeability assay (PAMPA) and ALogP values, among other measurements. Besides being able to enter cells more easily, each molecule's degradation activity also improved after this otherwise ``small'' change to the linker. 

While small-molecule drugs (SMDs) have demonstrated success in treating various diseases,\cite{xie2021small, dong2021discovery, dodson2022biologics} PROTACs can target and degrade proteins regardless of their function by hijacking the cell's natural disposal mechanism -- the ubiquitin-proteasome pathway. \cite{neklesa2017targeted} This approach circumvents the inherent limitations of standard SMDs, which must occupy specific binding sites on target proteins. \cite{Békés2022PROTAC}
Notably, $\sim$85\% of the human proteome has been deemed ``undruggable'' by SMDs;\cite{hopkins2002druggable, neklesa2017targeted, overington2006many, oprea2018unexplored} these proteins typically lack well-defined binding pockets, offering limited opportunities for ligand interaction or binding.\cite{Békés2022PROTAC, dang2017drugging} For instance, 63\% of the known 600 cancer-related proteins are classified as ``undruggable'', including transcription factors, scaffold proteins, and membrane-bound proteins. \cite{vicente2022mdm2} While SMDs are constrained to a limited pool of proteins that can be effectively targeted, PROTACs don't necessarily require binding to a specific well-defined pocket on a POI to trigger degradation; in theory, they can bind to any reachable region on a POI's surface that facilitates induced proximity between the POI and an E3 ligase, even in the case of low binding affinities with the POI. \cite{lai2017induced, han2019discovery} This flexibility in target engagement expands the scope of the proteome that can be drugged with this modality, and is a mechanism that can be particularly pertinent in cancer treatments, where target proteins often develop resistance to SMDs through mechanisms like genetic mutations, overexpression, or altered signaling pathways. PROTACs offer a potential alternative capable of overcoming these resistance barriers.\cite{sun2018protac, buhimschi2018targeting, sun2019degradation, brand2019homolog, jiang2019development}

Furthermore, a key advantage of PROTACs lies in their substoichiometric catalytic activity, which operates on an event-driven basis. \cite{bondeson2015catalytic} This means that PROTACs do not need to fully occupy their target proteins to be effective, in contrast to traditional inhibitors that function in an occupancy-driven manner. \cite{Konstantinidou2019PROTACs}
In SMDs, the effectiveness of the drug is often dependent on stoichiometrically occupying the target binding site. \cite{nath2009ubiquitin}  This means that a significant portion of a target protein must be bound by an inhibitor molecule for the desired therapeutic effect to be observed. This often requires relatively high concentrations of the drug to achieve sufficient occupancy, since the effects are proportional to the extent of binding. \cite{nath2009ubiquitin}  PROTACs operate differently: they bind transiently to their targets and, after facilitating ubiquitination, dissociate. This allows them to cycle through multiple rounds of activity, repeatedly initiating the degradation of additional instances of the same POI. \cite{lai2017induced} In contrast to SMDs that act in a dose-dependent manner, this catalytic feature allows PROTACs to achieve potent effects at possibly lower doses, offering potential advantages in terms of efficacy, safety, negative side effects, and off-target effects. \cite{olson2018pharmacological} 

One final advantageous characteristic of PROTACs worth mentioning is that they are able to selectively target and induce the degradation of specific protein isoforms. These are distinct forms of the same protein arising from a single gene. The ability to selectively target them is significant because it implies that PROTACs can be used to differentiate between closely related forms of a protein and target only the isoform(s) associated with a disease without affecting others that may have essential functions in normal cellular processes. \cite{bondeson2015catalytic, martin2021protacs, vicente2022mdm2}

These attractive characteristics make PROTACs a prime focus of drug design endeavors. To maximize the potential of this innovative class of compounds, researchers are increasingly turning to data-driven approaches for design strategies.  Machine learning (ML) has thus demonstrably advanced drug discovery and development by enhancing target identification, small-molecule design, predictive biomarker discovery, and the prediction of clinical trial success.\cite{vamathevan2019applications, dara2022machine} ML methods can help researchers analyze large amounts of data to identify potential drug targets, optimize compound properties, and predict how patients will respond to treatments. This makes drug development more efficient and increases the likelihood of success \textit{in vitro}. \cite{vamathevan2019applications, dara2022machine} Given the complexity of designing PROTACs due to the large chemical space they span and their multivalent nature, leveraging ML will likely be crucial in making the development of this new modality more feasible. Despite numerous reviews on PROTACs, there is a notable gap in the literature: an in-depth review that delves into the use of ML for PROTAC design is still lacking. In this comprehensive literature review, we explore the impact of ML on \textit{de novo} PROTAC design to date. First, we delve into the distinct characteristics of PROTAC linker design, underscoring the features required to create effective bifunctional molecules capable of TPD. We then examine how ML in the context of fragment-based drug discovery (FBDD; Figure \ref{fgr:FBDD}a), honed for small-molecule drug discovery, is paving the way for PROTAC linker design. Our review provides a critical assessment of the obstacles inherent in the application of these methods to PROTAC development. This assessment seeks to shed light on the pressing need for specialized algorithms, enhanced data quality, and the adaptation of ML models to address the multifaceted nature of PROTAC engineering. Moreover, we review existing ML works that have been tailored to PROTAC design, highlighting pioneering efforts as well as the limitations associated with these existing approaches. We also offer perspectives on potential avenues for future ML research in this field. 

\section{Machine learning in PROTAC linker design}

\subsection{The peculiarities of PROTAC linker design}

The role the linker plays in PROTAC function is both unique and complex, offering the broadest scope in terms of where structural modifications can be made when designing a PROTAC (Figure \ref{fgr:FBDD}b). \cite{zorba2018delineating, chan2018impact, nowak2018plasticity, smith2019differential, guenette2022target,cecchini2022linkers} Unlike the linker, the structures of the ligands that bind the POI and E3 ligase are generally more restricted. These restrictions stem from the need to maintain specific structure-activity relationships (SAR) and effective target binding, limiting the options for modifying the warhead and E3 ligase ligand according to their functional requirements. \cite{cecchini2022linkers} Consequently, the linker becomes a primary focus for design optimization in PROTACs. Modifications to the linker, such as altering its length, tuning its flexibility or rigidity, and incorporating different chemical groups, can influence the pharmacokinetic (PK) and pharmacodynamic (PD) profiles of PROTACs, as well as their degradation activity and overall efficacy.  \cite{cecchini2022linkers, bemis2021unraveling, bondeson2018lessons}  Notably, the geometry (conformation) of the ternary complex is heavily influenced by the nature of the linker. \cite{gadd2017structural, roy2019spr} Evidently, the linker not only dictates the spatial arrangement necessary for successful TPD in a given system but also the efficiency with which the PROTAC can facilitate the degradation of the POI. \cite{cyrus2011impact, bondeson2018lessons} This is often quantified by metrics like the \dc, the concentration at which half-maximal degradation is observed, and \dmax, the maximum level of degradation achievable. This implies that PROTAC linker design requires a multifaceted approach that balances several key properties to ensure the successful degradation of a POI while maintaining a desirable drug-like profile. In this section, we present example case studies that highlight the unique aspects of linker design in PROTACs.

\textbf{Linker length}. The conformation of the ternary complex is heavily dependent on the linker length within PROTACs.  A long linker can lead to the formation of no functional complex since the ubiquitination of the POI might not occur. A short linker might result in what's called a binary complex where the PROTAC is only effectively linked to either the POI or the E3 ligase, but not both. \cite{wang2016new} Similar results were observed by \citet{qin2018discovery}, where the potency of PROTACs targeting BET proteins to inhibit cell growth was highly dependent on the linker length. This was demonstrated via a series of five PROTACs with progressively longer linkers. Notably, the authors observed an optimal linker length; further extending the linker beyond this optimal length did not enhance the potency. This shows that a very long linker might not provide additional benefits, and could inadvertently introduce steric constraints leading to decreased binding affinity. The most effective PROTAC from this study, QCA570, was tested in xenograft mouse models of leukemia, where it induced complete and durable tumor regression at low picomolar concentrations.

Furthermore, the length of the linker affects the range of spatial configurations accessible to potential ternary complexes during formation, restricting which protein interfaces are accessible for interaction. \citet{smith2019differential} demonstrated how differences in linker lengths and attachment points enable selective degradation of closely related kinase isoforms using PROTACs. The study developed isoform-selective PROTACs for the p38 mitogen-activated protein kinase (MAPK) family using the same warhead and E3 ligase but varying the linker features (linker attachment points and lengths). Two different linker attachment points (an amide and phenyl series) and varying linker lengths (10, 11, 12, and 13 atoms) were used to create distinct PROTACs that differentially recruit VHL. This selective recruitment controls the degradation of either the p38$\alpha$ or p38$\delta$ isoforms. For instance, PROTACs with 12- and 13-atom linkers in the amide series became highly selective for p38$\alpha$ degradation, showing much higher degradation efficacy compared to degraders with shorter linkers which were also less selective. Conversely, a 10-atom linker in the phenyl series led to selective degradation of p38$\delta$, with very minimal impact on other isoforms. This selective degradation ability is attributed to how variations in linker lengths and attachment points influence the formation of the ternary complex. By fine-tuning the linkers, PROTACs can achieve selective degradation profiles -- in this particular study, shorter linkers may bring the E3 ligase into a position that is optimal for ubiquitinating p38$\delta$ but not p38$\alpha$.  

\begin{itemize}
    \item Linker length is an important factor in determining the spatial configuration necessary for effective ternary complex formation. Adequate length ensures optimal potency. Both too-long and too-short linkers can negatively impact the potency of PROTACs. 
    \item Small changes in linker length can shift the degradation selectivity between closely related protein isoforms.
\end{itemize}

\textbf{Linker composition}. The linker composition has an impact on the physicochemical properties of PROTACs, such as solubility and membrane permeability, among other factors. \cite{yokoo2023investigating} For instance, the substitution of amide bonds with ester bonds in BET degraders MZ1 and ARV-771 results in improvements to their permeability and cellular activity.\cite{klein_amide--ester_2021} The ester-linked versions, OMZ1 and OARV-771, demonstrated a 10- and 1.5-fold increase in PAMPA permeability, respectively. This improved permeability contributes to their enhanced ability to degrade target proteins, with OMZ1 achieving a 1.5- to 2-fold increase in degradation potency, while OARV-771 achieved a 5.5-fold increase. Despite concerns about the stability of esters compared to amides, both OMZ1 and OARV-771 maintained stability in plasma, indicating that their increased permeability does not compromise their overall stability. Additionally, an optimal lipophilicity range (ALogP between 3 and 5) was established for these ester-linked PROTACs, balancing effective membrane crossing with adequate aqueous solubility and minimal efflux. \citet{klein_amide--ester_2021} conclude that amide-to-ester substitution can benefit the optimization of PROTACs, and potentially other compounds, falling beyond the Rule of 5.

The composition of the linker can also improve other physicochemical properties of PROTACs, such as metabolic stability, and biodistribution. \cite{cecchini2022linkers} These properties influence how the drug is adsorbed, distributed, and eventually metabolized inside the body. 
However, the majority of linkers in PROTACs have been based on a limited set of chemical motifs, with PEG and alkyl chains being the most common. Approximately 55\% of linkers utilize PEG, while about 30\% use alkyl chains of various lengths. \cite{troup2020current} These motifs are favored due to their versatility, ease of synthesis, and ability to modulate the solubility and permeability of PROTAC molecules. Around 65\% of published PROTAC structures incorporate both alkyl and PEG segments within their linkers. This combination aims to leverage the beneficial properties of both motifs, such as the flexibility and hydrophilicity provided by PEG, and the structural simplicity and modifiability of alkyl chains. A further 15\% of linkers involve modifications to the basic glycol units in PEG, such as adding methylene groups. \cite{troup2020current} Such modifications are typically done to explore different chain lengths and thus influence the potential structural configurations accessible to PROTACs.

\begin{itemize}
    \item Amide-to-ester substitution can benefit the optimization of PROTACs, and potentially other compounds, falling beyond the Rule of Five.
    \item Modifications in PROTAC linker composition, such as altering chemical groups and combining different motifs, directly influence the physicochemical properties of PROTACs.
\end{itemize}

\textbf{Linker flexibility}. The flexibility of the linker can allow a PROTAC to more easily adapt to different spatial configurations, though too much flexibility may also lead to less predictable interactions.\cite{nowak2018plasticity, crew2018identification} A certain degree of rigidity can thus confer stability during ternary complex formation, leading to more consistent degradation activity. A 2018 study on PROTACs for ANK-binding kinase 1 (TBK1) degradation underscores the role linker flexibility plays in PROTAC activity.\cite{crew2018identification} Flexibility was imparted via long alkyl and ether chains to achieve potent compounds due to the inherent flexibility of these chemical units. A systematic exploration of linker length was thus conducted, and the degradation activity of each PROTAC variant was measured, focusing on those with sub-µM potency to identify effective linker lengths and compositions. The study found that PROTACs with linkers shorter than 12 atoms showed no appreciable degradation activity. In contrast, longer linkers, despite their higher polar surface area and challenges in cellular penetration, were generally well-tolerated and effective in degrading TBK1. The very flexible nature of the linker allowed long linkers to orient the ligands in a way that facilitated the association of TBK1 and VHL into suitable ternary complexes. 

Another study describes how the ability of PROTACs to induce selective protein degradation is enhanced by the plastic nature of the binding interactions between CRBN and BRD4 bromodomains. \cite{nowak2018plasticity} \textit{Plasticity} here means that the proteins can adopt multiple conformations at the binding interface depending on the linker length, composition, and linkage position. It was shown that different linkers can promote different binding conformations between the CRBN and BRD4. This plasticity allows the PROTAC to effectively bring the proteins into proximity in orientations that are conducive to ubiquitination. Using X-ray crystallography and molecular docking, the authors shed light on how different linker configurations lead to distinct low-energy binding conformations between CRBN and BRD4. The varying conformations accessible to PROTACs in this study illustrate how linker-induced flexibility directly impacts biological outcomes.

\begin{itemize}
    \item The flexibility of a PROTAC's linker can be tuned by adjusting, for instance, the length and chemical composition of the linker. 
    \item Flexibility can allow for conformational adaptability and access to multiple binding orientations.
\end{itemize}  

\textbf{Bottom line} -- There are no definitive guidelines that guarantee the design of an effective PROTAC for any given E3 ligase-POI pair. \cite{troup2020current} This means that developing a potent degrader generally requires trial and error, with a reliance on empirical metrics to identify effective linker features that establish the optimal SAR. The large combinatorial space makes linker design ideally suited to data-driven approaches, which provide a valuable complement to traditional, labor-intensive experimental methods. Computational models can furthermore simulate how ternary complexes form and behave at a level of detail generally inaccessible in most experiments. Used wisely, computational tools can aid in tricky tasks such as linker optimization without the need to synthesize and test numerous variants experimentally, potentially speeding up their development and reducing the costs associated with the synthesis and empirical testing of unpromising compounds.

One recent study which nicely illustrates these points used a combination of crystallographic data and mathematical modeling to explore the conformational dynamics of protein-protein interactions induced by PROTACs, to understand how these dynamics influence ubiquitination and eventual protein degradation.\cite{zhao2024structural} Interestingly, the authors found that the stability of the ternary complex did not necessarily correlate with increased protein degradation efficiency, suggesting that excessive stability might inhibit degradation efficiency. Notably, the spatial arrangement and kinetic properties of the ternary complex were crucial in this context: effective PROTACs brought lysine residues on the POI close to the active site of the E2 enzyme, facilitated by the E3 ligase within the complex. Lysine residues are the most common sites for ubiquitination in proteins. The authors also confirmed that the kinetics of the ternary complex, especially its dissociation rate, also play a role in determining the degradation efficiency. Salt bridges and the hydrophobicity of the interactions within the ternary complex were found to contribute positively both to the cooperativity and to the half-life of the interaction. These findings suggest prioritizing compounds that can induce the necessary conformational dynamics without overly stabilizing the ternary complex, highlighting how valuable insights can be gained using computational tools.

\subsection{PROTAC linker design goes beyond fragment linking}
\begin{figure*}[h!]
 \centering
 \includegraphics[scale=0.6]{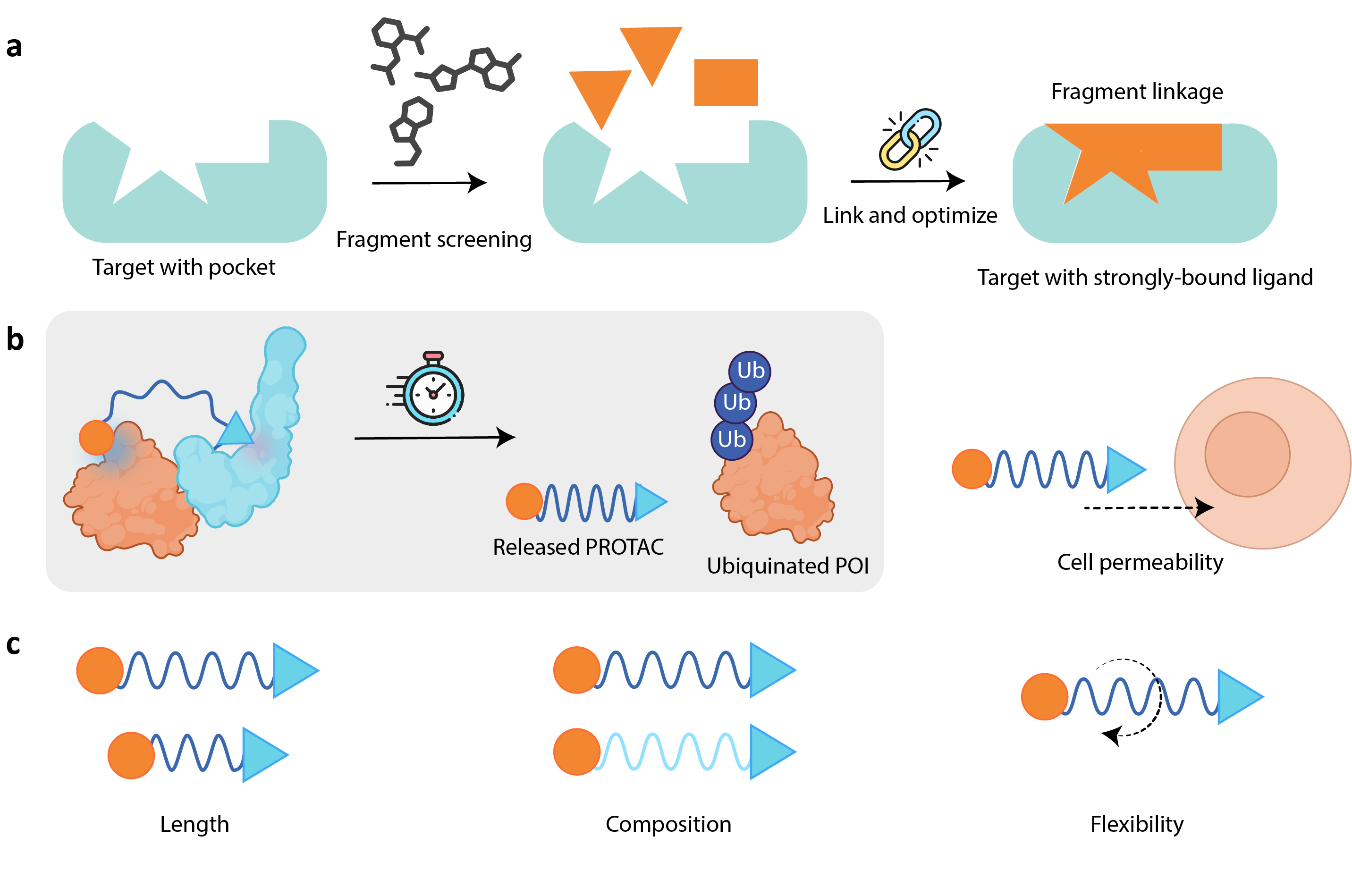}
 \caption{(a) An overview of fragment-based drug design (FBDD). The initial step involves fragment screening to identify potential fragments that can bind to the pocket of the target protein. These fragments are then linked and optimized to improve their binding properties. The result is a strongly-bound ligand that fits precisely within the target protein's pocket. (b) \textit{left} -- The linker in a PROTAC isn't just a passive bridge. It's an important component that enhances the interaction dynamics between the POI and the E3. \textit{right} -- The linker also contributes to the PROTAC's overall PK profile, including cell permeability. \textit{center} -- Because its MoA relies on transient ternary complex formation, the PROTAC is eventually released, meaning it is catalytic and can go on to be reused for other processes inside the cell. (c) The large and multivalent nature of PROTACs means they require a more complex design approach than FBDD methods developed for small molecules. The linker must be long and/or flexible enough to allow the warhead and E3 ligase ligand to adopt the necessary conformations for effective ternary complex formation, but not too flexible that the PROTAC cannot maintain the correct spatial orientation of the warhead and E3 ligase ligand. The linker may also need to incorporate specific chemical groups to enhance the overall potency of the PROTAC.}
 \label{fgr:FBDD}
\end{figure*}

While PROTAC design shares similarities with traditional small-molecule drug design, it is fundamentally distinct, notably in linker optimization (Figure \ref{fgr:FBDD}c). For instance, the approach to optimizing PROTAC linkers differs significantly from the concept of ``fragment linking'' used in fragment-based drug design (FBDD). In essence, FBDD is a strategy used in drug discovery where small, low-complexity molecules, i.e., fragments, are screened for binding to a specific pocket on the target protein. First, a library of small chemical fragments is created. These fragments are typically smaller than traditional drug-like molecules, with a molecular weight of less than 300 Da. The fragment library is then screened against the target protein to identify fragments that bind to the protein. This can be done using various techniques such as nuclear magnetic resonance (NMR) spectroscopy, X-ray crystallography, surface plasmon resonance (SPR), and thermal shift assays. Fragments that show binding affinity to the target protein are identified as ``hits.'' These hits often bind with low affinity but serve as a starting point for further optimization. Subsequently, the identified fragments are optimized to improve their binding affinity and drug-like properties. This can involve growing the fragment by adding more atoms, merging fragments that bind to adjacent sites, or linking fragments that bind to different parts of the target protein. Finally, the optimized fragments are developed into lead compounds, which have improved pharmacological properties and can be further tested in biological assays and in \textit{vivo} studies. \cite{bon2022fragment, li2020application} 

Fragments can be an ideal starting point for drug design, with fragment growing and linking strategies allowing for the optimization of their potency and physicochemical properties. Fragment linking in particular gives the possibility for significant potency gains by ensuring that the linked molecule maintains the interactions of the original fragments, a phenomenon known as super-additivity.\cite{guo2023link} However, achieving this is in practice very challenging, as a bad linker can instead lead to the disruption of fragment binding poses.

Despite its success in drug discovery, FBDD may fall short when applied to PROTAC linker design. PROTACs are substantially larger and more complex than the small fragments typically dealt with in FBDD. The linker in a PROTAC must connect two distinct binding moieties, facilitating the formation of a stable ternary complex, and does not simply focus on improving the binding affinity. To reiterate, the linker in a PROTAC must be flexible enough to allow the formation of a ternary complex but rigid enough to maintain the correct spatial arrangement of the ligands. This balance is difficult to tackle using traditional FBDD approaches, which focus on optimizing single-binding interactions rather than complex multi-protein assemblies. The unique challenges posed by the size, complexity, and spatial requirements of PROTACs necessitate more advanced methodologies. While direct application of typical fragment-linking strategies used in FBDD is not generally feasible in PROTAC design, a modular approach can certainly be beneficial. As we show in the next section, researchers are already taking inspiration and lessons learned from FBDD and applying them to PROTAC design.

\subsection{FBDD and ML pave the way for PROTAC development}

The advent of ML in drug discovery has transformed the way researchers approach the design and optimization of therapeutics. ML algorithms streamline labor-intensive design-make-test-analyze (DMTA) cycles by automating the identification of promising compounds and enabling researchers to avoid synthesizing and testing ineffective compounds, thereby reducing both time and costs.\cite{vamathevan2019applications, dara2022machine} Before the widespread use of ML in drug discovery seen today, computational tools for linker design typically involved searching a database, limiting proposed linkers to those already documented in the database.\cite{guo2023link} While such approaches have been successful, ML-based solutions can give models an advantage by allowing them to better explore the possible chemical space. Generative models, in particular, are used to create new molecules from scratch, designing possible structures with enhanced or desirable property profiles based on patterns learned from data. \cite{cheng2021molecular, pang2023deep} However, generative models trained for FBDD are optimized for designing linkers between small fragments. These models might not account for the larger size and higher complexity of PROTAC structures, potentially leading to inaccurate linker design. Presented below is a detailed overview of generative models used in \textit{de novo} linker design within the framework of FBDD.

\textbf{2D generative models in linker design}. SyntaLinker\cite{yang2020syntalinker} is a method that employs syntactic pattern recognition with molecular string representations via deep conditional transformer neural networks.\cite{yang2020syntalinker} It uses molecular representations based on the simplified molecular-input line-entry system (SMILES), one of the most popular methods for representing 2D molecular graphs as strings. SyntaLinker allows for the automatic linking of molecular fragments by learning from patterns in drug-like compounds in ChEMBL.\cite{gaulton2012chembl} This enables the generation of new molecular structures starting from pairs of fragments in a conditional way; possible constraints include the shortest-linker bond distance (SLBD), HBDs, number of rotatable bonds, and number of rings. The constraints are incorporated as control codes. For example, ``[L\_4]'' denotes a linker with a bond distance of four, which acts as a guiding prefix in the sequence. ChEMBL data was filtered using Lipinski's ``Rule of Five,'' PAINS substructures, and synthetic accessibility (SAscore) to ensure lead-like and synthesizable molecules were used to construct the training data. The dataset was constructed using matched-molecular pairs (MMPs) to build molecules into fragment molecule triplets (fragment 1, linker, fragment 2). SyntaLinker consists of multiple encoder–decoder stacks where each encoder layer has a multi-head self-attention sub-layer and a feedforward network sub-layer. The model was trained using the prepared dataset, with SLBD and the aforementioned constraints to learn the implicit rules of fragment linking. The generated molecules were evaluated using several metrics, including validity (97.2\%), uniqueness (88.1\%), recovery (84.7\%), and novelty (91.8\%).

Link-INVENT\cite{guo2023link} is an extension to the existing \textit{de novo} molecular design platform REINVENT; it uses policy-based reinforcement learning (RL) for multi-parameter optimization, and can be applied to both fragment linking and scaffold hopping given a desired property profile. Via RL, the Link-INVENT agent learns to generate linkers connecting molecular fragments while satisfying diverse objectives, facilitating the practical application of the model for real-world drug discovery projects. In the original study, Link-INVENT used the drug-like compound SMILES extracted from ChEMBL for training. Lenient criteria were applied to ensure the dataset's effectiveness for PROTAC applications (e.g., larger warheads). ChEMBL compounds were sliced using reaction SMIRKS to create triplets (linker, warheads, full molecule). Unrealistic data points were removed, and datasets were augmented via SMILES randomization for improved generalizability. Link-INVENT is trained based on the conditional probabilities of observing a linker given both molecular subunits, similar to SyntaLinker. The agent is initialized with the same parameters as the prior and is updated via RL to generate linkers that increasingly satisfy the desired multi-parameter optimization (MPO) objectives. The scoring function combines various components (physicochemical properties, structural features, predictive models, and binding energy approximations) to evaluate the desirability of generated linkers. Link-INVENT was tested in various experiments, demonstrating its capability to generate linkers that meet specific criteria. Notably, Link-INVENT has also been demonstrated to be effective in PROTAC linker design, successfully optimizing the properties of generated linkers, including effective length, the presence of rings, and flexibility.

Due partly to the surprising effectiveness of 2D representations like SMILES, the majority of molecular generative models used for \textit{de novo} molecular design and FBDD have made limited use of 3D structural information, including SyntaLinker and Link-INVENT. Nevertheless, the PROTAC MoA suggests that incorporating 3D information may come to play an important role in designing PROTAC structures which lead to favorable ternary complexes. In the next subsection, we cover ML models that seek to incorporate structural information into their molecular design workflows.

\textbf{3D generative models in linker design}. 
DeLinker\cite{imrie2020deep} is a graph-based deep generative model that incorporates 3D structural information for designing molecules using a multi-modal encoder-decoder. Training data was derived from a 250k molecule subset of ZINC, better known as ZINC-250k.\cite{gomez2018automatic} The dataset was further processed to create fragment-molecule pairs using standard transformations from MMP analysis. 3D conformers were generated using RDKit, and the lowest-energy conformation was used as the reference structure.
Molecules are represented as graphs, where atoms are nodes and bonds are edges. The model builds new molecules iteratively, bond by bond, from a pool of atoms that can be initialized with partial structures. The model uses a gated graph neural network (GGNN) for encoding and a single-layer neural network for edge prediction and labeling. This setup allows the model to use local, global, and 3D structural information for molecule generation, encoding multi-modal molecular information in a low-dimensional latent space. The model is trained under the VAE framework to reconstruct known linkers from fragment pairs and the linked molecule. The latent vector is derived from the embedding of the linked molecule, and the model is regularized to follow a standard normal distribution. The training objective includes a reconstruction loss and a Kullback-Leibler (KL) regularization term, which together ensure that generated molecules are valid and structurally sound. DeLinker can generate novel molecular structures, including those with longer linkers of at least five atoms. It was applied to the design of PROTACs targeting SMARCA2 and SMARCA4 subunits, where it generated linkers maintaining the geometry observed in a ternary complex with VHL. \citet{imrie2021deep} further improves DeLinker in their more recent DEVELOP framework by introducing a convolutional neural network (CNN) which operates on the 3D structure of the starting fragments. DEVELOP not only improves the proportion of generated molecules with high 3D similarity to the reference molecule, but also recovers 10× more of the original molecules compared to DeLinker.

3DLinker\cite{huang20223dlinker} is a conditional generative model for designing molecular linkers using 3D spatial information and is capable of generating linker graphs along with their 3D structures and anchor atoms. This is achieved through an E(3)-equivariant graph VAE, addressing challenges such as the conditional generation of linkers based on two input ligands and the requirement for 3D structural awareness to avoid atom clashes. It predicts both the graph (2D) structure of the linker and its 3D coordinates while ensuring the model's outputs are equivariant with respect to E(3) group symmetries (i.e., rotation, translation, and reflection). The training data was derived from the ZINC database,\cite{irwin2020zinc20} from which 3D conformers were generated for each molecule using RDKit and the lowest-energy conformation chosen as the reference structure. The final curated dataset contains $\sim$366k (fragment, linker, coordinate) triplets and was roughly divided into 99.8\%/0.1\%/0.1\%  training/validation/testing splits. 
Using this generous training split, the model outperforms other baselines, including DeLinker and other 2D graph generative models (coupled with ConfVAE\cite{xu2021end} for 3D structure generation), in recovering molecular graphs and accurately predicting the 3D coordinates of atoms. Nevertheless, it is unclear if the reported metrics are for the training, validation, or test set. While 3DLinker demonstrates improved performance in generating 3D molecular structures with accurate geometry, precise connection of molecular fragments, and higher recovery rates, the authors observed that this comes with the trade-off of lower uniqueness and novelty in sampled molecules compared to the benchmarked approaches.

DiffLinker\cite{igashov2024equivariant} is an E(3)-equivariant 3D-conditional diffusion model for the design of molecular linkers. This approach uniquely generates molecular linkers for a set of input fragments represented as 3D atomic point clouds, overcoming the limitations of previous methods by not being restricted to linking pairs of fragments. DiffLinker automatically determines the number of atoms in the linker and its attachment points to the input fragments. As the previous approaches, DiffLinker was trained and evaluated on a dataset derived from ZINC-250k, but the authors also took things a step further by benchmarking on two additional datasets: one derived from CASF-2016, and another derived from GEOM\cite{axelrod2022geom}. The molecules derived from GEOM can be decomposed into three or more fragments with one or two linkers  connecting them, creating a more challenging benchmark that better approximates real-world usage. DiffLinker demonstrates an ability to generate diverse and synthetically accessible molecules with minimal clashes, especially when conditioned on target protein pockets. It represents a significant advancement in FBDD, providing a powerful tool for the generation of chemically relevant molecules in a flexible and efficient manner. Nevertheless, the authors did not apply their fragment-linking approach to PROTAC design.

Building upon the success of SyntaLinker, DRlinker\cite{tan2022drlinker} is a similar approach that incorporates RL, and, indirectly, 3D information, for the generation of linkers with specific 2D and 3D attributes. It was trained and evaluated for FBDD on datasets derived not only from ChEMBL, but also from CASF-2016. \cite{su2018comparative} On tasks like optimizing bioactivity, it achieves a 91.0\% and 93.9\% success rate in generating compounds with desired linker length and LogP, respectively. Despite being based on 2D SMILES representation, DRlinker can also perform scaffold-hopping in a way that generates molecules with high 3D similarity but low 2D similarity to lead inhibitors. Two years later, the same team followed up with another model for FBDD which aims to better incorporate 3D information. GRELinker\cite{zhang2024grelinker} combines a gated-graph neural network (GGNN\cite{li2015gated}) with RL and curriculum learning (CL) to design linkers with desirable property profiles. Its architecture is very similar to that of GraphINVENT.\cite{mercado2021graph} It outperforms DRlinker in tasks such as controlling LogP, optimizing synthesizability and bioactivity, and generating molecules with high 3D similarity but low 2D similarity to lead compounds. It has also been evaluated in scenarios representative of real-world use-cases, where the aim is to optimize for molecular affinity using docking scores. The authors found that the use of CL improved its efficiency in generating complex linkers.

Despite the successes of the aforementioned works in FBDD, and, in particular, of DeLinker and Link-INVENT in PROTAC linker design, the methods reviewed above all face a key limitation -- they were all trained and optimized on small-molecule binders rather than on an actual PROTAC dataset. Although careful filtering was done to make the datasets more generalizable beyond small-molecule binders, due to the fact that warheads can be much larger than the typical fragments used in FBDD, we argue that the training of these models may not fully capture the unique features and complexities of larger, multivalent molecules like PROTACs, nor their unique chemistry. As previously discussed, PROTACs not only have larger sizes but also exhibit different biophysical and chemical properties compared to the small molecules typically found in drug discovery databases (Figure \ref{fgr:descp_distribution}). This training limitation can affect the applicability of these methods for designing effective PROTAC linkers, as the chemical space and design strategies for PROTACs diverge significantly from those of small molecules. This underscores the necessity for specialized tools for PROTAC linker design that can accommodate their unique size, complexity, and 3D structural requirements. The next section reviews ML models specifically tailored for PROTAC design.

\subsection{Previous work in PROTAC linker design}

In this section, we focus on prior ML studies that have been designed to advance the optimization of PROTAC linkers. By highlighting how computational models have been intentionally developed to refine PROTAC linker design, we hope to illustrate how the application of these technologies has evolved from traditional drug design to TPD. As ML techniques continue to evolve, they are expected to play an increasingly central role in PROTAC development. Notable ML-driven works such as AIMLinker\cite{kao2023fragment} and PROTAC-RL\cite{zheng2022accelerated} shed light on the complex nature of PROTAC linker design, though they are not the only recent works to tackle this problem.

PROTAC-RL is a deep generative model combining an augmented transformer architecture with memory-assisted RL capable of generating PROTACs with favorable PK properties, including solubility, stability, and bioavailability.\cite{zheng2022accelerated} Notably, the authors experimentally validated their model by testing the synthetic feasibility of six of their designs.
To address the challenge of limited training data, the model was pre-trained using a large dataset of PROTAC-like structures, termed \textit{quasi-PROTACs}, followed by fine-tuning on actual PROTAC data. Given a pair of E3 ligand and warhead SMILES, the model generates optimized linkers, which aim to optimize the PK attributes of the returned PROTACs. PROTAC-RL achieved a recovery rate of 43.0\%, much higher than the recovery rates of the baseline models, DeLinker and SyntaLinker, even after these were retrained using the PROTAC training datasets. After retraining, Delinker and SyntaLinker achieved recovery rates of 4.8\% and 10.4\%, respectively. This stark contrast in recovery rates between PROTAC-RL and the benchmarked models after retraining further strengthens the argument that models designed and trained for small molecular fragments cannot adequately capture the unique aspects of PROTACs, as the design strategies and principles for these two classes of molecules are fundamentally different.
Because the RL component allows for the conditional generation of PROTACs with specific properties, such as a desired protein target, the authors applied PROTAC-RL to the design of BRD4-targeting PROTACs. To this end, they generated 5k compounds which they then filtered through a combination of ML classifiers and molecular simulations to identify candidates with favorable PK properties and synthetic accessibility. Of the six candidate PROTACs which were synthesized and experimentally tested, three showed inhibitory activity against BRD4 in cell-based assays. One lead candidate demonstrated high anti-proliferative potency and a favorable PK profile in mice.

AIMLinker is a GGNN\cite{li2015gated} model for autoregressive PROTAC linker generation at the atomic/bond level.\cite{kao2023fragment} Like GRELinker, it seeks to improve upon previously-developed graph-based deep generative models like DeLinker, CGVAE\cite{liu2018constrained}, and GraphINVENT\cite{mercado2021graph} via the incorporation of 3D information. AIMLinker was trained on a dataset combining molecules from ZINC and PROTAC-DB\cite{weng2023protac}. The training focused on predicting viable 2D linker structures from fragment-molecule pairs. Generated molecules were then validated via molecular docking and simulations to verify binding to the target proteins via binding affinity and conformational predictions. AIMLinker was used to successfully generate a diverse library of novel PROTACs. The model demonstrated superiority over other fragment-linking methods (DeLinker and DiffLinker) in generating molecules with favorable PK properties and high binding affinities, with a few designed PROTACs even outperforming the reference compound dBET6 in binding affinity and structural alignment (Figure \ref{fgr:MoA}c). Despite a promising performance in PROTAC linker design, AIMLinker does have two current limitations, namely the focus on a single PROTAC target (BRD4) and the reliance on docking predictions, which are known to be inaccurate. \cite{pantsar2018binding}

Finally, ShapeLinker\cite{neeser2023reinforcement} is a model based on Link-INVENT, but with an important shape alignment contribution to the scoring function, and less significant but still important contributions from the ratio of rotatable bonds and the linker length ratio. The authors train on PROTAC-DB\cite{weng2023protac} data, as well as on ten well-known ternary complexes from the Protein Data Bank (PDB): 5T35, 7ZNT, 6BN7, 6BOY, 6HAY, 6HAX, 7S4E, 7JTP, 7Q2J, and 7JTO. All of these complexes have binding PROTACs that were optimized in individual structure-based drug studies and cover a diverse range of PROTAC (and linker) ``shapes.'' These are included in the training of the shape alignment model. Nevertheless, it is not clear whether these additions indeed improve the performance of ShapeLinker over that of the base Link-INVENT. The results suggest that perhaps larger changes to the model architecture are required for step-changes in performance.

\section{Machine learning in \textit{de novo} PROTAC design: going beyond linker optimization}
\subsection{Comprehensive PROTAC design strategies}

The design of a PROTAC involves more than the design of the linker; it also includes the optimization of the ligands that bind to the target protein and the E3 ligase. The challenge lies in designing a molecule where these components work in harmony to achieve ubiquitination and subsequent degradation of the target protein. 
The design process begins with the selection of a ligand, if known, that can selectively bind to the POI. This step is crucial because the efficacy of a PROTAC largely depends on the ability of this warhead to recognize and attach to the intended POI. \cite{li2020proteolysis} 
When numerous known binders exist for a protein, the binding affinity, physicochemical properties, and synthetic feasibility of these binders are crucial factors to consider. Approved drugs, drug candidates in clinical trials, or highly active inhibitors may be preferred starting points due to their optimized PK and PD properties, for example. \cite{he2022strategies} 

Equally important is the choice of an E3 ligase and a corresponding ligand. The selection of the E3 ligase often involves considering a range of factors, such as the ubiquitination efficiency, the specificity of the ligase, and its expression levels within the relevant cells or tissues. For instance, if a PROTAC is being developed for cancer therapy, the chosen E3 ligase should be highly expressed in cancer cells and less so in healthy cells to minimize off-target effects. \cite{li2022proteolysis} Furthermore, different E3 ligases can induce varying degrees of degradation even with the same POI ligands and linkers. The selection of an effective E3 ligase and ligand is thus a critical aspect of PROTAC design, and structural knowledge of the E3 ligase-POI interaction can significantly aid this process. Interestingly, neither the binding affinity of the warhead nor of the E3 ligase ligand seems to directly influence the degradation efficiency of the PROTAC. \cite{he2022strategies}

Once the individual ligands have been selected, the next challenge is to design a molecule where these components work in concert. This harmony is essential for the formation of an effective ternary complex between the PROTAC, the target protein, and the E3 ligase. The spatial and temporal dynamics of this complex formation are critical. It's not just about bringing these entities into proximity; it's also about ensuring that they interact in a manner that facilitates the transfer of ubiquitin from the E3 ligase to the target protein.
For instance, the spatial arrangement in a potential ternary complex needs to allow the POI’s ubiquitination site to be accessible to the E3 ligase once the complex is formed. This may involve tweaking the linker length, rigidity, or chemical composition to achieve the optimal orientation.

Nevertheless, some ML-driven methods for PROTAC design seek to tackle the problem in a more holistic manner. Though less common than modular approaches which focus heavily on PROTAC linker optimization, comprehensive PROTAC design strategies can be advantageous for a few reasons. They allow, in principle, for the simultaneous optimization of multiple parameters, such as flexibility, cell permeability, and degradation efficiency, factors which are determined not only by the linker composition, but also by that of the warhead and E3 ligand. A holistic approach may also better account for the complex interactions between the different PROTAC components, leading to the design of more effective and specific PROTACs; nevertheless, this remains to be rigorously demonstrated. In the next section, we examine the only study which, to our knowledge, has tackled the problem of engineering PROTACs in a holistic fashion, challenging traditional FBDD principles.

\subsection{Previous work in comprehensive PROTAC design}
A case study in ML-driven \textit{de novo} PROTAC design is the application of GraphINVENT, a graph-based deep generative model, to the generation of novel PROTAC graphs predicted to represent highly active degraders.\cite{nori2022novo} The authors used policy-gradient reinforcement learning (RL) and a surrogate model for protein degradation activity to guide the model toward a chemical space of more active potential degraders. The open-source PROTAC-DB\cite{weng2023protac} database was used to train the model, which included 638 ``complete'' entries detailing degradation activity in various systems. For an entry to be considered complete, it needed to include a PROTAC SMILES, an E3 ligase, a POI, a defined cell type, and, crucially, a \dc \ value. The authors reported that, despite their large size, GraphINVENT did not struggle to propose novel PROTAC-like structures starting from empty graphs; they go on to show that, following RL fine-tuning, the model could generate diverse molecules with not only higher predicted activity than the prior, but also substructures found in known degraders. An analysis was conducted to demonstrate that the model could be used to generate novel compounds with high predicted activity for IRAK3 degradation, but none of the proposed structures were experimentally validated.

\subsection{Modeling degradation activity in PROTACs}
A critical component of any generative model for PROTACs is a reliable surrogate model for degradation activity. To this end, a few data-driven approaches have been developed to tackle the prediction of degradation activity in PROTACs. \citet{nori2022novo} first used eXtreme Gradient Boosting (XGBoost) and Morgan fingerprints to classify potential degraders into ``active'' or ``inactive'' compounds. Following this work, \citet{li2022deepprotacs} introduced DeepPROTACs, a deep neural network architecture integrating graph convolutional networks (GCNs) and bidirectional long short-term memory (LSTM) layers for predicting degradation activity in PROTACs. DeepPROTACs was trained using data from PROTAC-DB and other public sources, including 2832 labeled datasets split into 988 ``good'' degraders and 1844 ``bad'' degraders based on their \dc \ and \dmax \ values. When tested on a dataset of 16 PROTACs targeting ER and VHL, the model achieved a prediction accuracy of 68.75\%. For other PROTAC targets like EZH2, STAT3, eIF4E, and FLT3, the accuracy rates ranged from 65\% to 80\%. More recently, \citet{ribes2024modeling} introduced a neural network ensemble model for the classification of PROTACs into ``active'' and ``inactive'' compounds. Here, the authors combined data from PROTAC-DB and PROTACpedia\cite{protacpedia}, where only entries which had both a \dc \ and \dmax \ value reported were used. PROTACs were represented using Morgan fingerprints, while target proteins, E3 ligases, and cell types are each embedded into feature vectors using linear layers, and normalized. The model demonstrated superior performance to previous models for degradation activity classification, reaching a top test accuracy of 82.6\% when using a stratified data split, and a test accuracy of 61\% on unseen POIs. The authors conclude that the model will generalize well to novel PROTAC structures so long as both the POI and E3 ligase have been seen before in training.

\subsection{Innovations and emerging trends}
The field of PROTAC design is rapidly evolving along with new ML approaches to molecular design. One common thread in many of the aforementioned methods is the use of RL to learn optimal policies for PROTAC design through trial and error. Another emerging trend is the integration of 3D information into generative approaches. This allows for a more holistic view of the interactions between proteins and can lead to more effective PROTAC designs. Additionally, there is a growing trend towards the use of transfer learning, where a model developed for one task is reused as the starting point for a model on a second task. This is particularly useful in PROTAC design where the limited amount of public data poses a challenge.

Although diffusion models have not yet been applied to PROTAC design (only FBDD, as in DiffLinker\cite{igashov2024equivariant}), we believe they present a promising direction in PROTAC engineering, both for linker-only and holistic design strategies. Firstly, diffusion models excel at generating high-quality molecular structures by gradually transforming simple distributions into complex data distributions. \cite{yang2023diffusion} Secondly, diffusion models can naturally integrate 3D information, which allows for the design of PROTACs that account for the spatial arrangement and interactions between the POI, the PROTAC, and the E3 ligase. Diffusion models are also known for their robustness in handling noisy data,\cite{guo2024diffusion} and they can be integrated with existing generative and predictive frameworks in an online setting. \cite{uehara2024feedback} These capabilities of diffusion models make them natural choices to explore further for generating diverse and novel PROTAC structures.

For a detailed summary of all models surveyed in this work, please see Table \ref{tab:model_comparison}.

\begin{table*}
    \small
    \caption{Summary of previous work on ML-guided PROTAC design. SMILES is a molecular string representation constructed from 2D graphs. RL: reinforcement learning. LSTM: long short-term memory, a class of recurrent neural network. JT-VAE: the junction-tree variational autoencoder.\cite{jin2018learning} CNN: convolutional neural network. GGNN: gated-graph neural network.\cite{li2015gated} CL: curriculum learning. GCNs: graph convolutional networks. MLP: multi-layer perceptron. The annotation ``(w/ 3D coords)'' indicates that 3D information was used indirectly during structure generation (e.g., scoring), whereas ``+3D coords'' indicates 3D information was directly used in structure generation.
    }
    \label{tab:model_comparison}
    \begin{tabular*}{\textwidth}{@{\extracolsep{\fill}}lllll}
        Model & Year & Data & Type & Focus \\
        \hline
        SyntaLinker\cite{yang2020syntalinker}     & 2020 & SMILES                     & transformers             & fragment linking \\ 
        PROTAC-RL\cite{zheng2022accelerated}      & 2022 & SMILES                     & transformers+RL          & fragment linking \& PROTAC linker design \\
        Link-INVENT\cite{guo2023link}             & 2023 & SMILES                     & LSTM+RL                  & fragment linking \& PROTAC linker design\\  
        ShapeLinker\cite{neeser2023reinforcement} & 2023 & SMILES (w/ 3D coords)      & Link-INVENT              & PROTAC linker design \\
        DRlinker\cite{tan2022drlinker}            & 2022 & SMILES (w/ 3D coords)      & transformers+RL          & fragment linking \\
        DeLinker\cite{imrie2020deep}              & 2020 & 2D graphs (w/ 3D coords)   & JT-VAE                   & fragment linking \\
        DEVELOP\cite{imrie2021deep}               & 2021 & 2D graphs+3D coords        & JT-VAE+CNN               & fragment linking \\  
        3DLinker\cite{huang20223dlinker}          & 2022 & 2D graphs+3D coords        & E(3) eq. graph VAE       & fragment linking \\
        \citet{nori2022novo}                      & 2022 & 2D graphs                  & GraphINVENT\cite{mercado2021graph} (GGNN+RL)    & full (``holistic'') PROTAC design\\
        AIMLinker\cite{kao2023fragment}           & 2023 & 2D graphs (w/ 3D coords)   & GGNN                     & PROTAC linker design \\
        GRELinker\cite{zhang2024grelinker}        & 2024 & 2D graphs (w/ 3D coords)   & GGNN+RL+CL               & fragment linking \\
        DiffLinker\cite{igashov2024equivariant}   & 2024 & 2D graphs+3D coords        & 2D GNN+E(3) eq. 3D diffusion & linker size prediction \& fragment linking \\
        \citet{nori2022novo}                      & 2022 & Morgan fingerprints        & XGBoost                  & degradation activity prediction\\
        DeepPROTACs\cite{li2022deepprotacs}       & 2022 & SMILES+3D graphs           & GCNs+LSTMs               & degradation activity prediction \\
        \citet{ribes2024modeling}                 & 2024 & Morgan fingerprints        & MLP & degradation activity prediction \\
    \end{tabular*}
\end{table*}

\subsection{Datasets}
There are currently two main sources of openly-accessible, structured PROTAC data: PROTAC-DB and PROTACpedia.

PROTAC-DB is a public database designed to support the research and development of PROTACs.\cite{weng2023protac} It offers an online repository of structural and experimental data related to these molecules. Data in the database is manually extracted from the literature or calculated using specific programs. In the second release, the number of PROTACs was expanded to 3,270 and featured $\sim$360 warheads, $\sim$1,500 linkers, and $\sim$80 E3 ligands. As of June 2024, PROTAC-DB contains 5,388 entries. It also includes ternary complex structures for PROTACs.  
PROTAC-DB covers key aspects of PROTAC activity, including degradation capacity, quantified by metrics like \dc \ and \dmax; binding affinities between PROTACs (or PROTAC ligands) and target proteins and E3 ligases; cellular activities such as IC$_{50}$, EC$_{50}$, GI$_{50}$, and GR$_{50}$; and PAMPA and Caco-2 permeability data. Nevertheless, entries are not necessarily complete and there is a lot of missing data in PROTAC-DB, often because the original source does not report all aforementioned metrics.\cite{ribes2024modeling}

PROTACpedia is a curated database focused on PROTACs, containing detailed entries on 1,190 PROTAC molecules as of the latest update (October 2022).\cite{protacpedia} It contains high-quality data that has been carefully curated by experts, including information on $\sim$202 warheads, $\sim$65 E3 ligands, and $\sim$806 linkers. This platform facilitates the sharing and dissemination of critical PROTAC-related data to help expand PROTACpedia. Its collaborative nature encourages contributions to ensure that the database remains an up-to-date resource for researchers exploring PROTACs. 

There is a significant overlap in the activity distributions of structures deposited in PROTACpedia with those in PROTAC-DB. As of June 2024, there are 807 PROTACs present in both databases, identified via string comparison following canonicalization of PROTAC SMILES from both databases. In other words, roughly 68\% of PROTACs in PROTACpedia and 25\% of PROTACs in PROTAC-DB are present in both databases. We did not explore what fraction of the duplicate PROTAC structures correspond to duplicate entries between the two databases, as it is possible that a PROTAC may be present in both databases but still contain information about different sets of experiments.

As PROTACs represent a relatively new therapeutic modality, there is a relative scarcity in the number of publicly available crystal structures, especially for ternary complexes. Structures that are available in the PDB have most frequently been determined using cryogenic electron microscopy (cryo-EM), a technology that has revolutionized the field of protein structural biology. Nevertheless, because the PROTAC MoA is not particularly well-understood, generalizations are being made across a range of PROTACs based on limited mechanistic data. Researchers would be wise to exercise caution when generalizing too far beyond the scope of their models or experiments.

Part of the challenge in ML-driven PROTAC engineering stems from the limited amount of \textit{structured} data available. While public databases such as these have been greatly influential thus far in driving the development of ML tools for PROTAC design, without more comprehensive datasets, data-driven models will only be able to access a fraction of the vast chemical space accessible with PROTACs. Data scarcity becomes even more of a concern when considering factors like bioactivity, PK properties, and 3D structure in PROTACs. Low-data and low-resource learning can provide valuable strategies in the current scarce data landscape \cite{van2024deep}, but, ultimately, more high-quality, structured data will need to be systematically generated and deposited following FAIR data-sharing principles for researchers to truly harness the powers of ML in PROTAC design. We hope that, just as ML has become an invaluable tool for identifying hits and optimizing leads in small-molecule drug discovery pipelines, it will also transform the current paradigm of PROTAC engineering, making us wonder how we ever managed without it.

\section{Discussion}
\begin{figure*}[h!]
 \centering
 \includegraphics[scale=0.4]{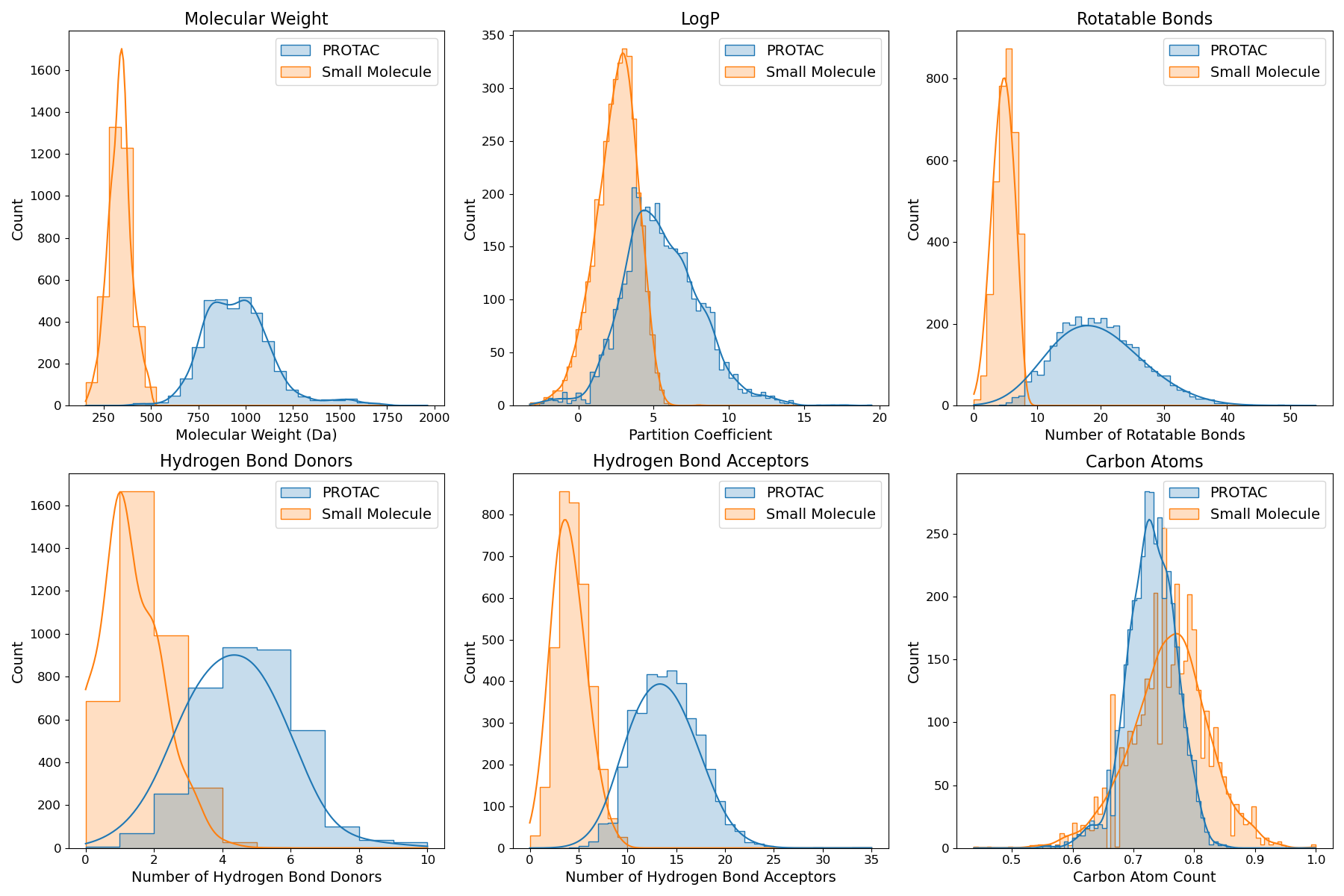}
 \caption{The distributions of various molecular descriptors in PROTACs versus small molecules. PROTACs were downloaded from PROTAC-DB and PROTACpedia, while small molecules were randomly sampled from ZINC-250k\cite{irwin2020zinc20}, a popular database used in drug discovery containing commercially-available compounds for virtual screening (e.g., drug-like compounds). This comparative analysis of their chemical and physical properties highlights the differences between both classes of molecules. The descriptors include molecular weight, partition coefficient (LogP), number of rotatable bonds, number of hydrogen bond donors (HBDs) and acceptors (HBAs), and normalized atom counts for carbon.}
 \label{fgr:descp_distribution}
\end{figure*}

ML models trained on small-molecule datasets often struggle with generalization to novel chemical spaces not represented in the training data. \cite{glavatskikh2019dataset, kretschmer2023small} This can limit their predictive accuracy for entirely new classes of compounds. 
PROTACs, with their bifunctional nature and larger size, represent a significant departure from small molecules. PROTACs generally have larger molecular weights and require higher degrees of conformational flexibility to achieve their function. Notably, the physicochemical properties and PK profiles of PROTACs are markedly different from those of traditional small molecules. In Figure \ref{fgr:descp_distribution}, we highlight some of the key physicochemical differences between PROTACs and SMDs. These include differences in molecular weight (MW), partition coefficient (LogP), number of rotatable bonds (flexibility and conformational dynamics), number of HBDs and HBAs, and number of carbon atoms. By analyzing these descriptors, we can gain precise insights into the structural and physicochemical differences between PROTACs and small molecules. 

\begin{itemize}
  \item \textbf{Molecular weight}. The small-molecule MW distribution peaks around 250-500 Da. This is the typical range expected for drug-like small molecules as it is considered optimal for oral bioavailability according to Lipinski's Rule of Five. The PROTAC distribution peaks around 750-1000 Da, highlighting how much larger they are than traditional small molecules. The small-molecule distribution is relatively narrow and sharply peaked, indicating a more uniform range of MWs, while the PROTAC distribution is broader, reflecting greater variability in the size of these complex molecules. The clear separation between the two distributions highlights a key difference between PROTACs and traditional small molecules: their size.

  \item \textbf{Partition coefficient}. The LogP distribution for small molecules peaks around 2-3. This is consistent with drug-likeness criteria, where a LogP value between 1 and 3 is typically considered favorable for oral bioavailability. The LogP distribution for PROTACs is broader and peaks around 5. Higher LogP values indicate that PROTACs are generally more hydrophobic than small molecules, which can affect their solubility and cellular permeability. The higher LogP values for PROTACs may pose challenges for their solubility in aqueous environments like the extracelullar environment and the cytosol, and may require formulation strategies to enhance their solubility and bioavailability.

  \item \textbf{Rotatable bonds}. The distribution in the number of rotatable bonds peaks around 1-5 rotatable bonds for small molecules. Fewer rotatable bonds are associated with greater rigidity. For PROTACs, the distribution instead peaks around 15-20 rotatable bonds. The higher number of rotatable bonds can be largely attributed to the flexible linker regions in PROTACs.

  \item \textbf{Hydrogen bonds}. The HBD distribution of small molecules peaks around 0-2. This is in line with the drug-likeness criteria that suggest a limited number of hydrogen bond donors to ensure good membrane permeability. On the other hand, the HBD distribution of PROTACs peaks around 3-5. Similar trends are observed for the HBA distributions.

  \item \textbf{Carbon composition}. Both PROTACs and small molecules have a high normalized carbon count, peaking between 0.7-0.8, with the peak being slightly lower for PROTACs. Small molecules display a slightly broader distribution in normalized carbon atom count. No significant differences in normalized nitrogen, oxygen, or fluorine atom composition were observed, although small molecules do display marginally broader distributions for all these atom types.
\end{itemize}

This comparative analysis highlights the unique challenges and opportunities facing ML models for PROTAC design. Furthermore, it should be evident that models trained on small fragments will not capture the distinct features of PROTACs, as small molecule fragments and PROTACs exist in largely non-overlapping areas of chemical space. Extending ML models developed for small molecules to PROTACs requires modifications; this could entail small changes, like re-training on larger molecules and/or more diverse datasets that include PROTACs, or changes to fundamental design principles. It is well-known in deep learning, including generative modeling, that the predictive accuracy of a model depends heavily on the availability of high-quality datasets.\cite{noauthor_data_2023} However, in drug design, datasets often have various forms of inconsistencies and missing data. \cite{myers2000handling} These challenges are even more pronounced when focusing on PROTAC design; here, experimental data on the efficacy and specificity of PROTAC molecules is even scarcer. The multifaceted nature of PROTACs necessitates detailed and high-quality datasets to uncover the subtle patterns underlying their biological activity. This is especially crucial for turning predictive models into tools for robust PROTAC optimization and design.

Models that fail to incorporate structural, or even dynamical, information regarding PROTACs and their target proteins might not effectively capture the feasibility of ternary complex formation. We know that the specificity and activity of PROTACs towards a specific POI are influenced by precise 3D interactions at the molecular level, more so than by the binding affinity. Without detailed 3D structural data, ML models may not be accurate enough to generalize to new PROTAC structures or even new POIs, a concern reflected in the changing landscape of ML models for PROTAC design: while early work focused primarily on 2D representations or simplified 3D information, all work we surveyed from the past two years involved the incorporation of more complex 3D data. We don't believe this change is due solely to advances in computing hardware and software. Rather, models capable of handling 3D data offer a superior capability to capture the interplay of molecular shapes and complex spatial arrangements especially relevant in PROTAC function.

Another big challenge with PROTAC design is getting them into the cell.\cite{yokoo2023investigating} As they depend on the proteasome for degrading their target proteins, PROTACs can only be used to target proteins found in the cytosol or with cytosolic domains (for membrane proteins), thus excluding as targets any proteins found outside the cell. According to the Human Protein Atlas, $\sim$25\% of all protein-coding human genes have been shown to encode proteins that localize to the cytosol and its substructures,\cite{humanproteinatlas} though this estimate does not include proteins that transiently reside in the cytosol. How to improve cellular permeability in PROTACs is thus an active area of research, as it imposes hard constraints on their efficacy. Notwithstanding, exactly which mechanism PROTACs use for entering the cell is not fully understood and may very well vary depending on the specific molecule and cell type, adding another layer of complexity to the task of cell permeability prediction.

To overcome the many challenges facing \textit{de novo} PROTAC design, future ML methods must place a greater emphasis on accurately modeling the 3D structures of PROTACs and the corresponding ternary complexes they form. This could involve the development of molecular dynamics or physics-based approaches that leverage ML to simulate important molecular interactions and conformational changes at a coarse-grained or even atomistic level, and it could also involve experimental advances that allow us to better isolate and characterize these complexes, possibly with the assistance of active learning or other ML-driven strategies. Scientists leveraging ML have undeniably driven many recent breakthroughs in protein structure prediction\cite{abramson2024accurate} and conditional protein structure generation.\cite{watson2023novo} Perhaps it's time to apply similar guiding principles to PROTAC engineering (e.g., systematic data collection and accessibility, better algorithms harnessing biological knowledge), and see what breakthroughs we can achieve in this domain.

\section{Conclusion}
PROTACs differ fundamentally from small-molecule drugs (SMDs) in their mechanism of action. SMDs such as inhibitors typically function by blocking a protein's active site and thus its activity, whereas PROTACs instead carry out a complicated dance inside the cell, which, if performed correctly, will lead to the degradation of the target protein. This means that rather than simply inhibiting a protein's function, a PROTAC removes it from the cell. Notably, it is not consumed in the process, which means it can go on to cause the degradation of many other copies of the target protein before it is metabolized and/or excreted. This can be more effective in cases where simple inhibition of activity is insufficient for a therapeutic effect, as is the case in many diseases lacking effective first-line treatments, including many cancers. PROTACs offer an alternative pathway to drugging otherwise ``undruggable'' proteins in the cytosol. 

In this comprehensive review, we have highlighted the significant impact of ML on PROTAC design. The complexities involved in PROTACs make traditional ML in the context of FBDD less effective. These complexities include the unique mechanism of action of PROTACs, the delicate spatial configuration required for effective protein degradation, and the need for favorable PK profiles for drug-like compounds, which are not adequately captured by models designed for small molecular fragments. Advanced ML techniques, such as generative models tailored to PROTAC peculiarities offer promising solutions for optimizing PROTAC design.

In the hope of spurring more research in what we view as a hugely impactful but formidable research direction, we have prepared this comprehensive review on ML for PROTAC design. We hope that this review and the insights described in it serve as a comprehensive guide to researchers looking to apply their deep ML knowledge to the design of an exciting ``new'' therapeutic modality, or conversely, to enable biologists to venture into the rewarding world of deep generative models. The synergy between ML and PROTAC design holds immense potential, and we encourage further research in this pioneering domain.


\section*{Author Contributions}
YG conducted the initial literature review, with guidance from RM. YG prepared the initial draft of the manuscript and figures. RM and YG engaged in extensive discussions and iterations of writing and feedback throughout the process. Both authors contributed significantly to the development and refinement of the manuscript. Both authors have read and approved the final version of the manuscript.

\section*{Conflicts of interest}
There are no conflicts to declare.

\section*{Acknowledgements}
YG and RM acknowledge the funding provided by the Wallenberg AI, Autonomous Systems, and Software Program (WASP), supported by the Knut and Alice Wallenberg Foundation. Some of the figures use icons from \href{https://www.biorender.com/}{BioRender} and \href{https://www.flaticon.com/}{Flaticon}.
\balance

\bibliography{mybibliography} 
\bibliographystyle{rsc} 

\end{document}